\documentclass[authoryear,preprint,review,12pt]{elsarticle}

\usepackage{amssymb}

\usepackage{hyperref}

\usepackage{lineno}

\journal{Ocean Dynamics}

\begin{document}

\begin{frontmatter}



\title{On the potential of mapping sea level anomalies from satellite altimetry with Random Forest Regression}


\author[DGFI]{Marcello Passaro}
\author[DGFI]{Marie-Christin Juhl}


\address[DGFI]{Deutsches Geod\"atisches Forschungsinstitut der Technischen Universit\"at M\"unchen, Arcisstra{\ss}e 21, 80333 Munich, Germany. Contacts: marcello.passaro@tum.de, +49 (89) 23031-1214}


\begin{abstract}
The sea level observations from satellite altimetry are characterised by a sparse spatial and temporal coverage. For this reason, along-track data are routinely interpolated into daily grids. The latter are strongly smoothed in time and space and are generated using an optimal interpolation routine requiring several pre-processing steps and covariance characterisation. In this study, we assess the potential of Random Forest Regression to estimate daily sea level anomalies. Along-track sea level data from 2004 are used to build a training dataset whose predictors are the neighbouring observations. The validation is based on the comparison against daily averages from tide gauges. The generated dataset is on average 10\% more correlated to the tide gauge records than the commonly used product from Copernicus. While the latter is more optimised for the detection of spatial mesoscales, we show how the methodology of this study has the potential to improve the characterisation of sea level variability.
\end{abstract}



\end{frontmatter}


\section{Introduction}  
The monitoring of sea level is conventionally performed using tide gauges and a network of radar altimeters orbiting the Earth. Tide gauges are in-situ data that register measurements at high frequency (often more measurements per hour) and are scattered irregularly along the global coastlines \citep{woodworth2016}. Altimeters sample along satellite tracks, spanning the same area after a defined number of days depending on the chosen repeating orbit \citep{fu2001}.

Since altimetry data are along-track measurements scattered in time and space, interpolating algorithms are routinely used to generate sea level maps that are regularly sampled in space and time. The European Union's Earth observation programme Copernicus currently releases daily sea level maps and their along-track sources through the Copernicus Marine Service (CMEMS).

The CMEMS daily maps are produced using a processing based on optimal interpolation, requiring several steps and assumptions described in \cite{letraon1998,taburet2019}. The along-track data are sub-sampled and filtered twice, using variable cut-off wavelengths ranging from 200 km to 65 km depending on the latitude. The optimal interpolation uses a variable number of observations in time and space, with spatial correlation scales ranging from 80 to 400 km and time correlation scales ranging from 10 to 45 days. It is based on the best linear least square estimator described by \cite{bretherton1976}, in which the covariance matrix of the observations is needed as an input. The covariance matrix is provided by means of assumptions on the errors of the different geophysical corrections applied to the along-track measurements \cite{pujol2016}.

It has been recently argued that data-driven interpolation is able to perform better than conventional optimal interpolation schemes, whose choice of covariance priors tend to over-smooth the sea level variability \citep{lguensat2019}. The concept behind data-driven interpolation is to exploit machine-learning to provide an estimation based on patterns and statistical relations acquired from the training data, rather than from external instructions and assumptions \citep[][]{zhou2017}.

The objective of this paper is to adapt an established machine learning technique to the problem of the estimation of daily sea level maps from along-track altimetry measurements. This is part of an effort aimed at finding new strategies to improve the characterisation of sea level variability at sub-seasonal time scales and in the coastal and shelf seas to "reduce the gap" between altimetric and tide gauge observations, as motivated by previous works such as \cite{cipollini2016}.


We use the Random Forest Regression algorithm, described by \cite{breiman2001}, in the implementation of \cite{pedregosa2011}, which has been already successfully used to fill gaps due to missing observations of the ocean (e.g. \cite{gregor2017empirical}). The particular choice for the Random Forest Regression is related to our adaptation of the scheme for spatial interpolation proposed by \cite{leirvik2021}, who tested several approaches with a similar aim and found best results using Random Forest Regression. 

In this study, the method is tested on a regional scale for one year of data in the North Sea and validated using tide gauge data and the optimally-interpolated maps from CMEMS. While CMEMS daily grids have only been validated using monthly averages from tide gauges as ground truth, we adopt in this work the daily averages of the Global Extreme Sea Level Analysis (GESLA, version 3), a global archive of high-frequency tide gauge data \citep{woodworth2016, haigh2021}.





\section{Data}

We consider the year 2004 and the extended North Sea including Skagerak/Kattegat in the east and the English Channel in the west.  in our case study. The year 2004 is chosen to limit the amount of data to be processed for this experiment, since the available altimetry mission in this year are restricted to: Jason-1, Envisat, Topex/Poseidon, Geosat Follow-on.

\begin{figure}[t]
\includegraphics[trim=7cm 0cm 0cm 0cm,width=18cm]{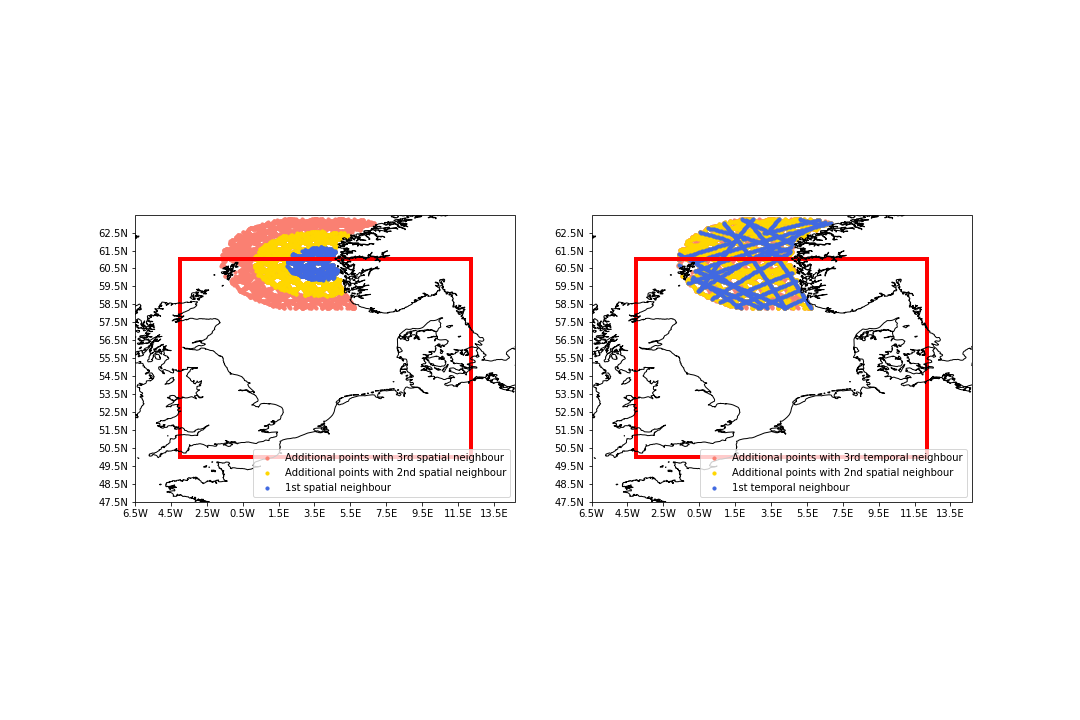}
\caption{Examples of along-track observations included in spatial (left) and temporal (right) neighbourhoods associated to one particular location. The red box indicates the area of study. The latter is extended in the search for neighbouring observations, in order to allow for the estimations at the domain's border.}
\label{fig_neighbourexample}
\end{figure}

The North Sea is chosen because of the extensive tide gauge network provided for validation and for the known challenges that gridded altimetry products encounter in terms of performances, as shown considering monthly data by \cite{dettmering2021} and \cite{woppelmann2016}. The region of interest and its geographical coordinates are delimited by the red square in Figure \ref{fig_neighbourexample}.

To train the Random Forest Regression, we use the CMEMS Level 3 (i.e.,along-track) sea level anomalies (SLA), reference number: \texttt{SEALEVEL\_\-GLO\_PHY\-\_L3\_REP\-\_OBSERVATIONS\-\_008\_062}. We recall that the SLA is defined as the sea level above the mean, corrected for atmospheric and tidal effects. A list of all applied corrections is available in \cite{taburet2019}.

We compare the daily machine-learning-based SLA from this study (nicknamed ML from now on) with the latest version of the CMEMS Level 4 gridded SLAs, reference number: \texttt{SEALEVEL\_GLO\_PHY\_L4\_MY\_008\_047}.

We stress the difference in the use of the data sources from CMEMS: the along-track data are the observations that are used to build the regression model; the gridded SLAs are only used for comparison with respect to the results of this study.

As external truth for the validation, we use high-frequency data from tide gauges available from the Global Extreme Sea Level Analysis (GESLA-3, \url{www.gesla.org}, \cite{woodworth2016}). In order to make the tide gauge data comparable to the altimetry dataset, the following processing steps are needed. Firstly, the atmospheric component is suppressed using the same correction applied to obtain the SLAs, i.e. the dynamic atmosphere correction from \cite{carrere2003}. Secondly, the tidal variability is suppressed using a 40-h LOESS filter, which has been tested to most effectively reduces tidal variance at periods lower than 2 days by \cite{saraceno2008}. Finally, the mean of the sea level record is computed and subtracted from each time series.



\section{Method}
The concept of our methodology is the use of along-track SLAs as truth to train the random forest regressor in the estimation of unknown SLAs (our target variable) on a set of grid points.

As predictors, we use means, weighted means and standard deviations of the SLAs at different neighbourhoods in space and time. Moreover, the ratios among these predictors from different neighbourhoods are also used as predictors, to better describe the evolution of the target variable in space and time.

This methodology is inherited from \cite{leirvik2021}, who used spatial neighbourhoods to constrain a Random Forest Regression for the interpolation of a surface solar radiation dataset. We expand the methodology by considering the time dimension as well.

The following subsections are dedicated to the details of our implementation.

\subsection{Preliminary steps}
All along-track data for 2004 from CMEMS are collected in the area of study, enlarged by 2.5 degrees in latitude and longitude to guarantee the definition of the neighbourhoods at its borders.

The target variable $y_{training}$ to train the regressor is the field \textit{sla\_unfiltered}, which consists in the SLAs at the posting rate of 1 Hz (roughly one measurement every 7 km along the track). The CMEMS Level 4 gridded SLA uses the field  \textit{sla\_filtered} when interpolating Level 3 data. Such field is a smoother version of the along-track data obtained using variable filter lenghts of several tens of kms. Our experiments have shown that the neighbourhood method proposed in this study does not need further filtering and our objective is to keep as much signal as possible. Further discussion and comparison with the CMEMS Level 4 with these regards is provided in Section \ref{Results}.

We define the locations where to compute the SLA, our unknown independent variable $y$, as the geographical coordinates of an daily unstructured grid spaced by 0.125 degrees in latitude and longitude, i.e. the same grid resolution of the CMEMS Level 4 product.

\subsection{Definition of neighbourhoods}

We define 3 spatial neighbourhoods and 3 temporal neighbourhoods to group the along-track altimetry observations in the proximity of the locations in $y_{training}$ and $y$.

The spatial neighbourhoods are concentric circles with a radius of 100 km, 200 km, and 300 km from the location of the target variable. The temporal neighbourhoods contain the along-track data collected within 5, 10 and 15 days from the time set by the target variable, within a distance that does not exceed 300 km. An example of the along-track locations assembled through the neighbourhoods of one target variable is provided in Figure \ref{fig_neighbourexample}. 

The borders of the neighbourhoods are selected to be within the average global correlation scales of sea level in time and space (see for example Figure 4 from \cite{pujol2016}). Nevertheless the choice for this experimental study is empirical and could be further optimised, for example by using global maps of variable correlation scales depending on the region, such as what is done in the generation of the CMEMS Level 4 grids. We anticipate that we do not observe a substantial change in the performances by slightly changing the neighbourhood definitions.

\subsection{Definition of predictors}

We define in this section the following classes of predictors: time and space clusters, single-neighbourhoods statistics and multi-neighbourhoods statistics.

\subsubsection*{Time clustering}

\begin{figure}[t]
\includegraphics[trim=5cm 0cm 0cm 0cm,width=15cm]{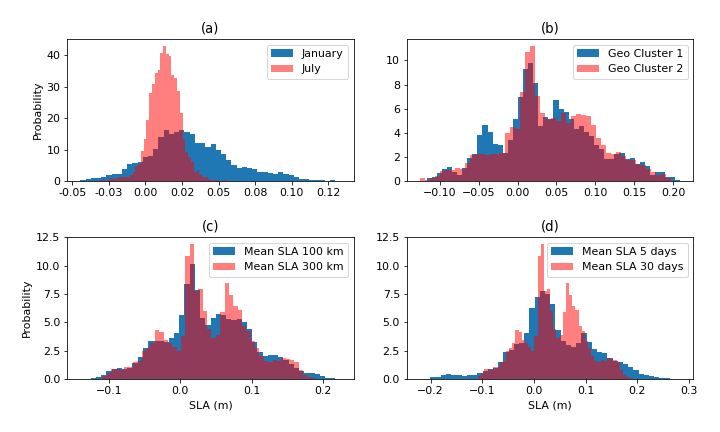}
\caption{Probability Density of the sea level anomalies associated to specific predictors from the training dataset. Panel (a): months of January and July. Panel (b): two geographical clusters. Panel (c): the mean of the sea level anomalies for the first (100-km radius) and third (300-km radius) spatial neighbourhoods. Panel (d): the mean of the sea level anomalies for the first (5-days) and third (30-days) time neighbourhoods.}
\label{fig_histograms}
\end{figure}

The time cluster contains the month in which the variable of interest is defined. Given that the annual cycle is the most prominent SLA periodic signal in time series whose length cannot catch decadal variability, we expect this information to be relevant for the regression. Indeed, Figure \ref{fig_histograms}a shows the two very different probability densities (PDs) of the SLA for January (blue) and July (red) based on the full training dataset.

\subsubsection*{Spatial clustering}

Several choices could be done concerning spatial clustering. In this exploratory study and in order to generalise the approach, we choose the agglomerative hierarchical clustering \citep{ward1963} in the implementation of \cite{pedregosa2011}. This is an unsupervised classification method that we use to separate the domain in different regions, simply based on the euclidean distance between the locations in our case. We choose to divide our subdomain in 9 clusters, an example of the different PDs of SLA from two of them is visualised in Figure \ref{fig_histograms}b. We reckon that this is a choice driven by simplicity and other oceanographic information could be used to refine the clustering, for example taking in consideration the spatial correlation with respect to tide gauges (the so called "zone of influence" approach from \cite{oelsmann2020}).

\subsubsection*{Single-neighbourhoods statistics}

For the SLAs contained in every spatial and temporal neighbourhood we compute the following statistics: mean, spatial-based weighted mean, time-based weighted mean and standard deviation. The weighted means are based on inverse distance weighting, i.e., maintaining the notation of \cite{leirvik2021}, the weighted means are defined as:

\begin{equation}
\tilde{z}(N) = \displaystyle\sum\limits_{z_i\in N} \lambda_i z_i
\end{equation}

where $N$ defines the neighbourhood, $z_i$ is every SLA values within it and the weights $\lambda_i$ are defined as:

\begin{equation}
\lambda_i = \frac{d_{i0}^{-r}}{\displaystyle\sum\limits_{z_i\in N}d_{i0}^{-r}}
\end{equation}

For the spatial-based weighted mean, $d_{i0}$ is the euclidean distance in km of every SLA within the neighbourhood and the location of the target variable. For the time-based weighted mean, $d_{i0}$ is the time difference in seconds between the passing time of the altimeter at the observation $i$ and the time stamp of the target variable. Note that the time difference is multiplied by a factor $10^{-4}$ in order to achieve similar orders of magnitude between spatial-based and time-based weighted means. 

The exponent $r$ expresses the relative importance of close-by observations. This study does not include a specific research on the optimisation of $r$, but we chose to keep $r=2$, as we found better performances than the choice of $r=5$ from \cite{leirvik2021}. We find our choice to be more representative of our problem, since a high exponential gives a high importance to the closest observations, while SLA is a field characterised by large spatial and temporal scales of correlation. 

Given 3 spatial neighbourhoods and 3 temporal neighbourhoods, we obtain therefore 24 single-neighbourhoods predictors. An example of the different PDs of the predictors is given in Figure \ref{fig_histograms}, where the PD of the mean SLA for the first and third spatial (panel c) and temporal (panel d) neighbourhoods is provided.


\subsubsection*{Multiple-neighbourhoods statistics}
The multiple-neighbourhoods statistics are the ratios between the single-neighbourhoods statistics of the same kind for consecutive neighbourhoods. For example, as in \cite{leirvik2021}, considering the mean of the SLAs we compute $\overline{Z}^{k1,k2}$ (ratio of the mean SLAs between first and second neighbourhoods) and $\overline{Z}^{k2,k3}$ (ratio of the mean SLAs between second and third neighbourhoods). Considering the typical objective of the altimetry missions to achieve a 1 cm SLA accuracy at 1 Hz posting rate \citep{bonnefond2012b}, we round up (or down, for negative numbers) to the cm the single neighbourhood statistics previously obtained before computing each ratio.

\subsection{Final steps}
The predictors are computed for both $y_{training}$ and $y$ locations, generating the predictor matrices $X_{training}$ and $X$, in which each row corresponds to the predictors associated with each location. Outliers in $y_{training}$ and $X_{training}$ are identified using a $3\sigma$ criterion, where $\sigma$ is the standard deviation of each variable. Observations in which the SLA or its predictors are identified as outliers are eliminated from the training dataset. Finally, the Random Forest Regression is applied on the training dataset. The obtained regressor $f(\cdot)$ is then applied to estimate the desired SLA on the grid points as $y = f(X)$.

\section{Results and Discussion}
\label{Results}

\subsection{Examples}
\begin{figure}[t]
\includegraphics[trim=5cm 0cm 0cm 2cm,width=15cm]{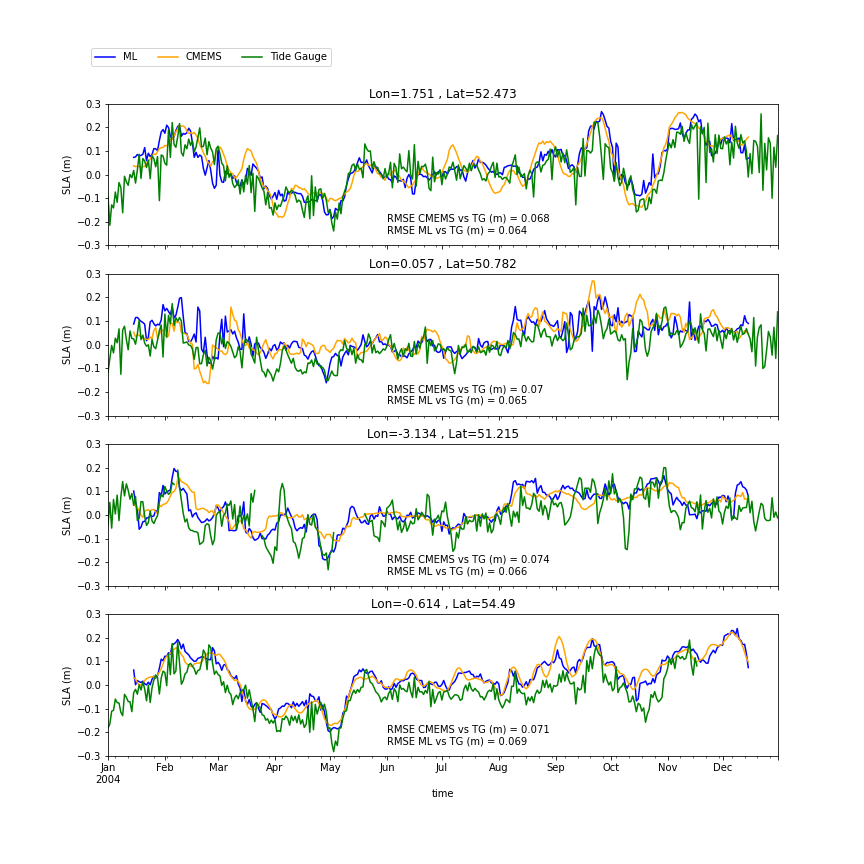}
\caption{Time series estimated from satellite altimetry from this study (ML, blue) and CMEMS (orange) at the closest point to four tide gauges (green), whose coordinates are shown at the top of each panel. Also shown as text is the Root Mean Square Error (RMSE) of the altimetry dataset considering the tide gauges as ground-truth.}
\label{fig_timeseriesexample}
\end{figure}

To investigate the advantages and the limitations of the generated daily ML product, we first consider examples in time and space. Figure \ref{fig_timeseriesexample} shows the time series of daily averaged data from tide gauges (in green), whose locations is specified at the top of each subplot. The ML product (in blue) and the CMEMS product (in orange), corresponding to the closest location to each tide gauge, are shown for a period comprised between the 15th of January and the 15th of December. This latest choice is due to the fact that we have only worked with data from 2004 and therefore the regression would generate worse results at the beginning and at the end of the period investigated.

The CMEMS time series appears significantly smooth in time, while ML preserves time scales that better match the ones of the tide gauges, although of course not the full extent of the high-frequency variability is captured. Despite CMEMS being smoother than ML, the root mean square error (RMSE), computed taken the tide gauges as the truth, is systematically lower for ML. This gives us confidence that the ML time series is not simply noisier than the CMEMS, but it is indeed more accurate.

\begin{figure}[t]
\includegraphics[trim=5cm 0cm 0cm 2cm,width=25cm]{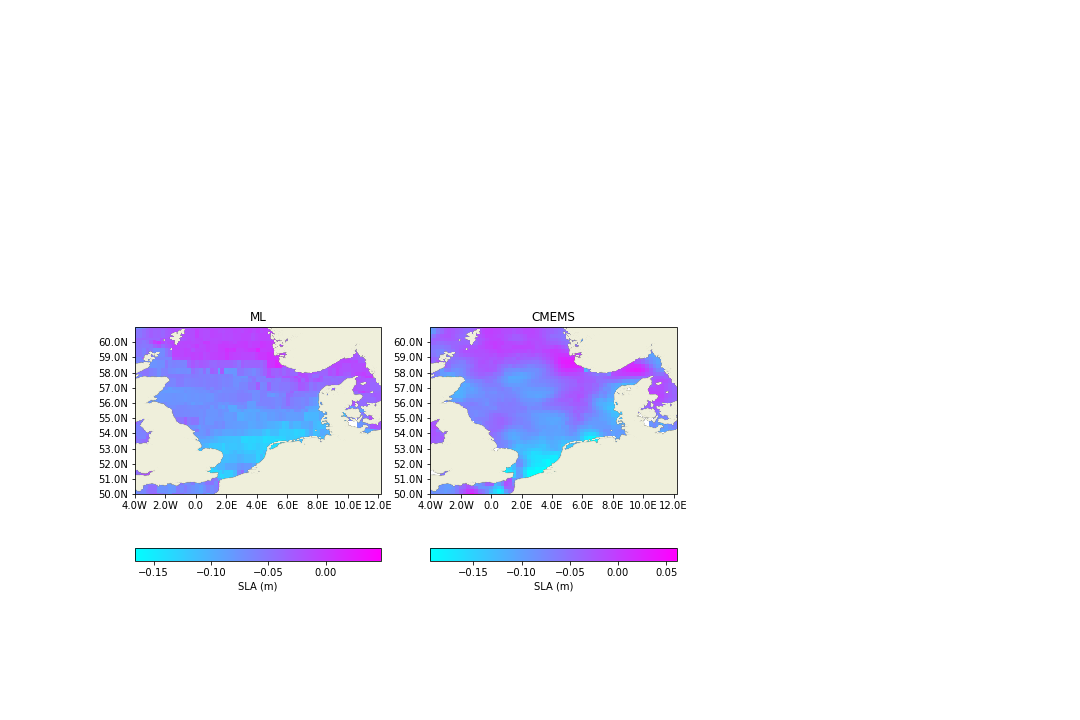}
\caption{The daily maps of sea level anomalies (SLA) from this study (ML) and CMEMS estimated for the 24th of April 2004.}
\label{fig_dailymapsexample}
\end{figure}

In Figure \ref{fig_dailymapsexample} we show a snapshot of ML and CMEMS SLAs for the 24th of April 2004. While the large-scale gradients are similar in both products, the CMEMS map has more defined contours identifying mesoscale variability. The higher variability of ML is the counterpart in space of what has been seen in time in the previous example. The objectives of ML and CMEMS are indeed different: the CMEMS optimal interpolation scheme is dedicated to the retrieval of mesoscale structures \citep{taburet2019}, while with ML we attempt to achieve a better compromise for sea level monitoring. The latest statement is quantified and verified for this case study in Sections \ref{Validation against tide gauges} and \ref{SLA variability}. Here, we further notice that the CMEMS map is affected by unrealistic extremes of SLA in single pixels in particularly challenging areas such as the English Channel. This is remarkable, considering that the input along-track data of ML and CMEMS are exactly the same, except for the along-track filtering applied by the latter.

\subsection{Validation against tide gauges}
\label{Validation against tide gauges}
We assess the general performances of ML and CMEMS computing the Pearson's correlation coefficient (CORR) and the Root Mean Square (RMS) between the time series obtained from altimetry and the daily means of the tide gauge data at the closest grid point.

\begin{figure}[t]
\includegraphics[trim=5cm 0cm 0cm 2cm,width=18cm]{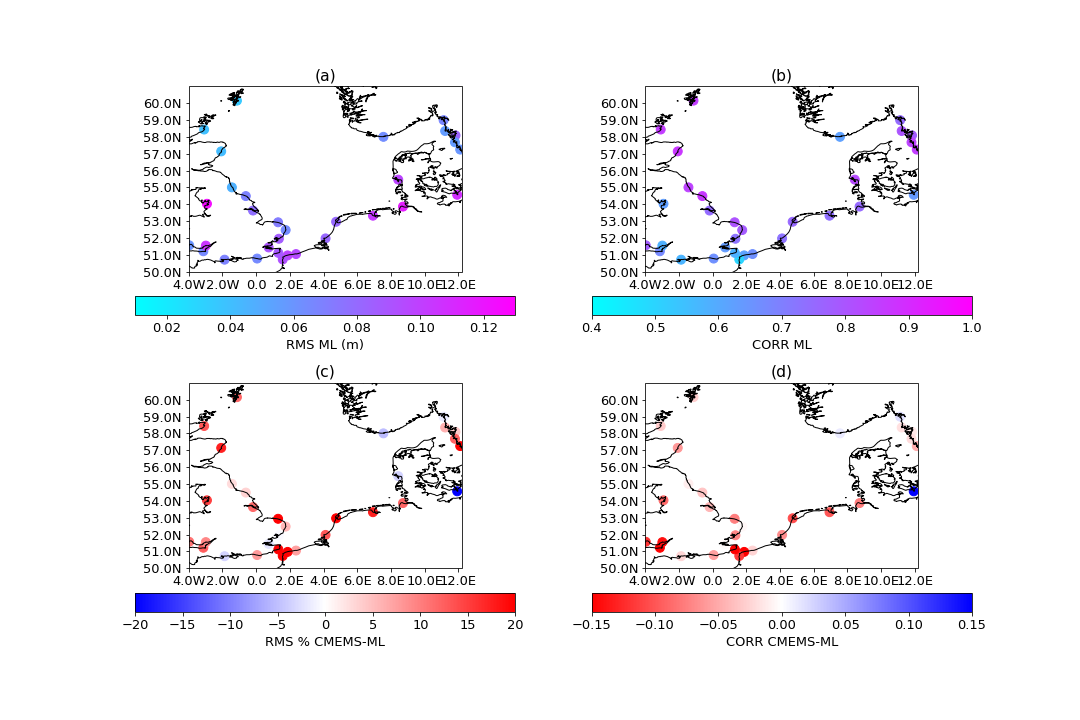}
\caption{Results of the validation of daily sea level anomaly maps coupled with tide gauges at the closest point. Root Mean Square (RMS, panel a) and Pearson's correlation coefficient (CORR, panel b) between the product of this study (ML) and the time series from the tide gauges (panel a). Panels c and d: Difference between these statistics and the equivalent computed using the CMEMS product, in which the red colour palette indicates an improvement using ML.}
\label{fig_corr_rms}
\end{figure}

Figure \ref{fig_corr_rms} shows in the upper panels the RMS and the CORR for ML and in the lower panels the difference with respect to the same statistics computed using CMEMS. The colorbar of the latter is adjusted in order to show towards the red every improvement brought by ML with respect to CMEMS.

Good performances (CORR$\>$0.7) are reached along the coasts facing a large open ocean area at the centre of the domain, such as the eastern coast of UK. Notably, good performances are also seen in much more enclosed areas situated at the periphery of the domain, such as the Kattegat Sea between Denmark and Sweden (easternmost part of the domain). This advocates for the robustness of the neighbourhood strategy previously presented. Lowest performances are reached in some enclosed bays and on both sides of the Channel between UK, France and Belgium (southernmost part of the domain). Here the quality of the SLAs, also in terms of the geophysical corrections used to extract them, plays a dominant role as shown in previous studies at different temporal scales, such as \cite{dettmering2021} using monthly time series.

The most remarkable result of the validation is that in almost all of the domain (29 tide gauges out of 32) ML performs better than CMEMS: In more than half of the domain there is at least a $5\%$ improvement in both CORR and RMS considering the tide gauges as ground truth.
The average improvement in correlation is 9.98$\%$ (6.99$\%$ considering RMS), with peaks over 30$\%$ that include some of the most problematic areas for satellite altimetry such as the Channel.

\subsection{SLA variability}
\label{SLA variability}
Finally, we assess how realistic the variability of the sea level from the daily grids is. For this purpose, we compute the interquantile range (IQR) of the time series at every grid point and at every tide gauge. The IQR is an index of variability computed by taking the difference of the 75th and the 25th percentile of the data, and it is typically used instead of standard deviation or variance because of its robustness. It is commonly used in sea level studies comparing in-situ and satellite time series (for example \cite{woppelmann2016}) and proves fit for our purposes, given that we only assess one year of data.

\begin{figure}[t]
\includegraphics[trim=5cm 0cm 0cm 2cm,width=18cm]{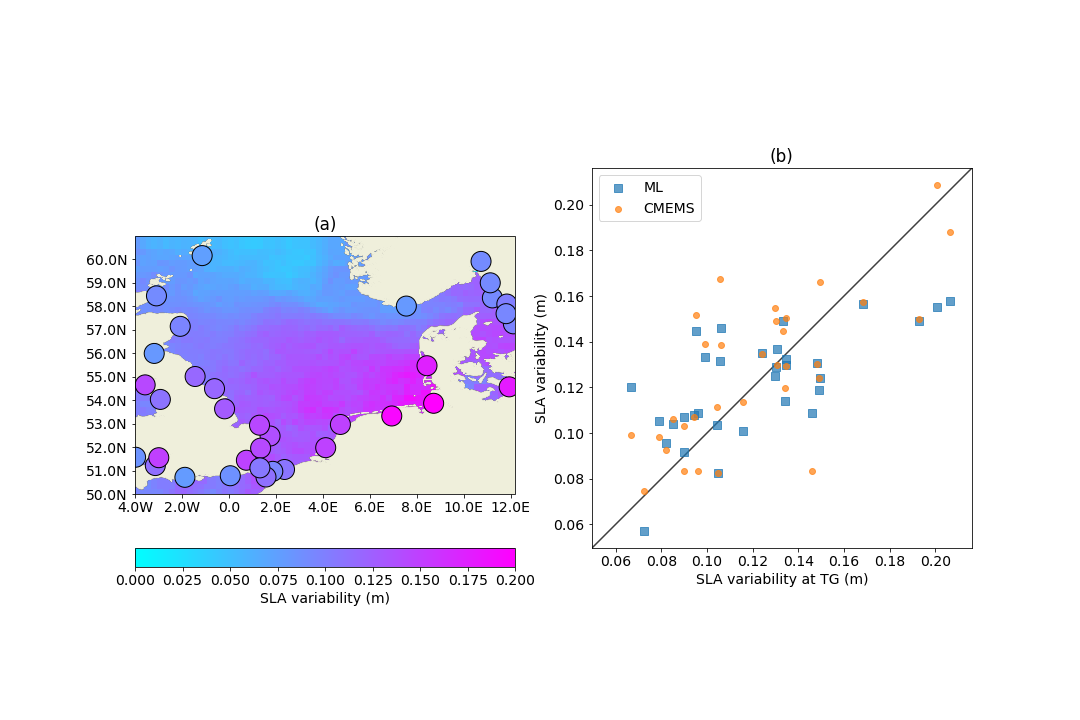}
\caption{Panel a: Variability of sea level anomaly (SLA) estimated using the interquantile range of the time series at each grid point estimated in this study (ML) and from the tide gauges (circles). Panel b compares the same statistics at the tide gauges (TG) with the closest grid point from ML and CMEMS products.}
\label{fig_variance}
\end{figure}

Figure \ref{fig_variance}a displays the results on the map, showing a consistent increasing variability of ML towards the south-eastern part of the domain, which is confirmed by the tide gauge records. In Figure \ref{fig_variance}b, the IQR at tide gauges is compared with the variability observed by ML and CMEMS at the closest point. To evaluate this comparison and considering the tide gauges as the ground truth, we compute an index of the average misrepresentation of the sea level variability:

\begin{equation}
    {Err}_{var}=\frac{\displaystyle\sum\limits_{i=1}^N  \frac{({IQR}_{alti} - {IQR}_{TG})}{{IQR}_{TG}}\cdot100  }{N}
\end{equation}

where $N$ is the number of tide gauges and ${IQR}_{alti}$ is the IQR of the altimetric time series at the closest point to each tide gauge. The best results are obtained by ML with ${Err}_{var}=4.4\%$, while using CMEMS ${Err}_{var}=7.6\%$ is achieved.

\section{Conclusions}  
This study has analysed the potential of using a data-driven approach to produce daily maps of SLAs starting from along-track observations from satellite altimeters. This approach allows for circumventing several hypothesis needed to characterise the covariance of the observations and their errors in the optimal interpolation. Building on the existing literature, we have tested a Random Forest Regression that uses statistics extracted from spatial and temporal neighbourhoods. By doing so, we have obtained one year of daily sea level maps that are on average 10$\%$ more correlated to the observations from tide gauge stations in the North Sea, compared to CMEMS data.

We believe that the main heritage of this study is the idea that along-track SLA data can be used to train machine learning routines aimed at generating gridded maps. The latter appear less smoothed in space than their CMEMS counterpart and will therefore need further filtering to be used for the identification of mesoscale features such as eddies. Nevertheless, the method presented allows for a more realistic representation of the sea level variability, as verified by the comparison against coastal in-situ data. Such comparison has been conducted using high-frequency tide gauges, which is in our opinion a much more realistic external validation than the use of monthly means, if the objective is to assess the capability of the altimetry constellation to observe sea level at short time scales.

Since this is an exploratory study, we have to acknowledge both potential and limitations. To speed up the experiments, we have chosen one single year of data (2004), in which 4 altimeters were in orbit, and a specific region (the North Sea). Extending this methodology to a longer time series will allow to perform coherence studies and distinguish therefore the performances at different time frequencies. We have used one single regressor, because clusters based on time and geographical locations of the observations were part of the predictors. Nevertheless, the feasibility of this choice will need to be assessed for studies involving more years and a wider area, also in terms of computing time.

The validation against tide gauges shows the strong potential of machine learning to improve the characterisation of coastal sea level at a time in which the altimetry community has recognised the possibilities to improve the quality of sea level data close to the coast \citep{benveniste2020}. We expect therefore further improvements by using SLAs whose estimation is optimised for the coastal zone \citep{passaro2021,birol2021}, which will nevertheless require significant post-processing of the along-track data, in order not to decrease the quality of the training dataset.


\section*{Description of author's responsibilities}
M.P. conceptualised and designed the study, is the author of the code used to generate and validate the results, and wrote the manuscript. M.-C.J. helped in the interpretation and discussion of the results, as well in the testing and improvement of the code. All authors read and commented on the final manuscript.

\section*{Acknowledgements}
The authors are grateful to Menghan Yuan, Julius Oelsmann, Luigi Cavaleri and Isabelle Puyol for their help, suggestions and interest.

\section*{Data References}
The dataset generated in this work is publicly available from: Passaro Marcello, Juhl Marie-Christin (2022). Daily sea level anomalies from satellite altimetry with Random Forest Regression. SEANOE. \url{https://doi.org/10.17882/89530}

The validation functions used to generate the statistics and plots in this work are publicly available from: \url{https://github.com/ne62rut/machine_learning_altimetry_validation}


\section*{Bibliography}
\bibliographystyle{plainnat} 





\end{document}